\documentclass[prd,twocolumn,superscriptaddress,floatfix,amsmath,amssymb]{revtex4-1}
\usepackage[english]{babel}
\usepackage{graphicx}
\usepackage{dcolumn}
\usepackage{bm}
\usepackage{verbatim}
\usepackage{mathrsfs}
\usepackage{cancel}
\usepackage{epstopdf}
\usepackage{color}
\usepackage{cancel}

\usepackage{float} 

\usepackage[usenames,dvipsnames]{pstricks}
\usepackage{epsfig}
\usepackage{pst-grad} 
\usepackage{pst-plot} 

\usepackage{hyperref}
\usepackage{subfigure}

\newcommand{\proj}[2]{\left| {#1} \right\rangle\!\left\langle {#2} \right|}

\newcommand{\ii}{\mathrm{i}}

\newcommand{\nn}{\nonumber}
\newcommand{\be}{\begin{equation}}
\newcommand{\ee}{\end{equation}}
\newcommand{\bea}{\begin{eqnarray}}
\newcommand{\eea}{\end{eqnarray}}

\def\slashchar#1{\setbox0=\hbox{$#1$} 
\dimen0=\wd0 
\setbox1=\hbox{/} \dimen1=\wd1 
\ifdim\dimen0>\dimen1 
\rlap{\hbox to \dimen0{\hfil/\hfil}} 
#1 
\else 
\rlap{\hbox to \dimen1{\hfil$#1$\hfil}} 
/ 
\fi}

\begin{document}

\title{Measuring motion through relativistic quantum effects}
\author{Aida Ahmadzadegan}
\affiliation{Department of Physics and Astronomy, University of Waterloo, Waterloo, Ontario N2L 3G1, Canada}
\author{Robert B. Mann}
\affiliation{Department of Physics and Astronomy, University of Waterloo, Waterloo, Ontario N2L 3G1, Canada}
\author{Eduardo Mart\'{i}n-Mart\'{i}nez}
\affiliation{Institute for Quantum Computing, University of Waterloo, Waterloo, Ontario, N2L 3G1, Canada}
\affiliation{Department of Applied Math, University of Waterloo, Waterloo, Ontario, N2L 3G1, Canada}
\affiliation{Perimeter Institute for Theoretical Physics, 31 Caroline St N, Waterloo, Ontario, N2L 2Y5, Canada}

\begin{abstract}

We show that the relativistic signatures on the transition probability of atoms moving through optical cavities are very sensitive to their spatial trajectory. This allows for the use of internal atomic degrees of freedom to measure small time-dependent perturbations in  the proper acceleration of an atomic probe, or in the relative alignment of a beam of atoms and a cavity.

\end{abstract}

\maketitle

\section{Introduction}\label{Intro}

Quantum metrology provides techniques to make precise measurements which are not possible with purely classical approaches. In quantum metrology protocols such as quantum-positioning and clock-synchronization \cite{Giovannetti2001,Giovannetti2006}, the exploitation of quantum effects such as quantum entanglement has allowed for a significant enhancement of the precision in estimating unknown parameters as compared to classical techniques \cite{reviewGiovannetti}. 

On the other hand, there exist metrology settings where general relativistic effects play an important role in establishing the ultimate accuracy of the measurement of physical parameters  \cite{AshbyGPS}. It is thus pertinent to introduce a framework where relativistic effects are  considered even in quantum metrology schemes \cite{RQMFuentes}, where it is relevant to study how (or if) incorporating relativistic approaches to quantum metrology may increase the precision and accuracy of the estimation and measurement of physical parameters. 

In this paper we focus on finding suitable quantum optical regimes where the response of particle detectors becomes sensitive to small variations of the parameters governing their motion, incorporating relativistic effects. Our goal is to assess the sensitivity of the response of particle detectors to such variations, in turn allowing for the precise measurement of such parameters.

In particular, we consider a setting in which an atomic detector crosses a stationary optical cavity while undergoing constant acceleration. Relativistic accelerating atoms in optical cavities have been considered before in the context of  an enhancement of Unruh-like radiation effect \cite{Scully,antiscully,rescully}, and later, in this context, to analyze the subtleties of the Unruh effect in the presence of boundary conditions \cite{Wilson2013}. The suitability of such settings as theoretical accelerometers was studied in \cite{Draganaccelerometer}, where it was shown that a detector's response is sensitive to variations of its proper acceleration.  
In this paper, we will show that near the relativistic regimes, but still, much below the accelerations required for the Unruh effect to be detectable, the detectors' response becomes sensitive to small (and maybe time-dependent) perturbations in either the parameters that govern their trajectory or in the alignment of the optical cavity. We will study this sensitivity to determine to what extent it is possible to exploit it for quantum metrological effects.

We consider two different scenarios of metrological interest.  In the first, we study the sensitivity of the response of the detector to  time-dependent variations of its proper acceleration. Specifically, we consider a uniformly accelerated atomic detector crossing an optical cavity with constant proper acceleration that undergoes a small harmonic time-dependent perturbation. If the system alignment is tuned, we might wonder how sensitive it is to the amplitude and frequency of the perturbation.

In the second scenario we study the sensitivity of the detector's response to variations of its trajectory. To accomplish this, we consider small harmonic perturbations of the spatial trajectory of a uniformly accelerated observer. We explore how sensitive this setting is to the amplitude and frequency of the perturbation, thus providing a setting to measure the wellness of the atom's trajectory alignment with respect to the cavity frame.

To this end, the outline of our  paper is as follows. In sec. \ref{set}, we introduce two physical settings including the methodology for investigating our two scenarios. Sec. \ref{result}, contains a discussion of our results.  Sec. \ref{concl} contains our concluding remarks.

\section{The setting}\label{set}

In this section we consider two different scenarios in which we want to precisely measure different parameters of the trajectory of an atomic probe. For the first scenario, which we will call the {\it accelerometer setting}, we consider an atomic probe following a constantly accelerated trajectory, but whose proper acceleration undergoes a harmonically time-dependent perturbation. In the second scenario, which we will refer to as the {\it alignment metrology setting}, we consider that the atomic probe's trajectory undergoes small harmonic perturbations as seen from the lab frame, so as to be able to measure the precision of the alignment of a cavity with a beam of atomic detectors.

In both scenarios we model the light-atom interaction by means of the Unruh-DeWitt model. Although simple, this model captures the fundamental features of the coupling between atomic electrons and the EM field involving no exchange of orbital angular momentum \cite{Wavepackets,Alvaro}. 
 
\subsection{A quantum accelerometer} 

Particle detectors with time dependent accelerations have been previously studied in \cite{Kothawala2010,Barbado2012a}, where the response of an Unruh-DeWitt detector with time dependent acceleration in the long time regimes has been considered in a flat spacetime with no boundary conditions. We would like to study how sensitive the detector response is to time-dependent perturbations of its proper accelerations in the short-time regime and in optical cavity settings.

In order to analyze this  accelerometer setting, let us first consider the parametrization of the trajectory of an atomic probe for a general time dependent trajectory in terms of the probe's proper time $\tau$  \cite{MollerRelativity}:
\bea\label{traj1}
x(\tau)&=&x_0+\int_{\tau_0}^{\tau} d\tau' \sinh \big[\xi(\tau')\big],\\
t(\tau)&=&t_0+\int_{\tau_0}^{\tau} d\tau' \cosh \big[\xi(\tau')\big],
\eea 
where 
\bea\label{speed}
\xi(\tau)=\xi_0+\int_{\tau_0}^{\tau} d\tau' a(\tau') 
\eea 
represents the atom's instantaneous speed, and $a(\tau)$ is the instantaneous proper acceleration of the probe.

For our purposes, we consider that the probe undergoes a constant acceleration, which is disturbed by a small harmonic perturbation:
\bea\label{four}
a(\tau)=a_0\big[1+ \epsilon \sin(\gamma \tau)\big]
\eea
$\epsilon$ and $\gamma$ are the respective relative amplitude and   frequency of the harmonic perturbation.

\begin{figure}[htp]
	\includegraphics[width=0.2\textwidth]{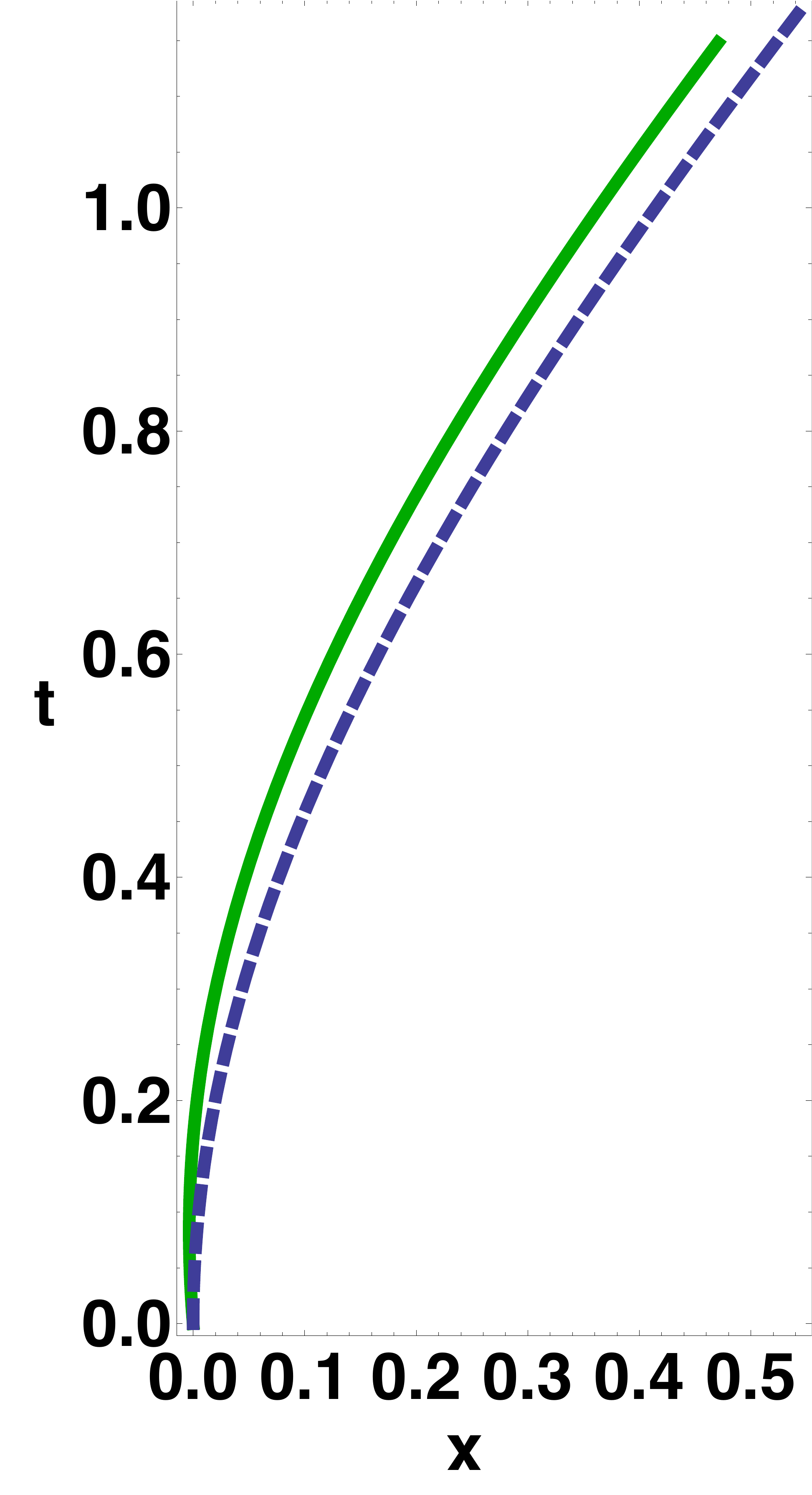}
  \caption{(Color online) The non-perturbed (blue-dashed curve) and perturbed (green-solid curve) trajectory for the accelerometer scenario. The trajectory is parameterized in terms of the proper time, $\tau$ of the detector.}
  \label{accelerometer}
	\end{figure}

The general form of the trajectories for both perturbed and constant accelerations is shown in Fig. \ref{accelerometer}.

In our setting, to find the transition probability of the detector, we let it cross a cavity of length L with an initial velocity $\xi_0$ and we measure its excitation probability for the period of time T that it spends traveling the full length of the cavity. The Hamiltonian that describes our system generates translations with respect to time $\tau$ in the detector's proper frame. This Hamiltonian consists of three terms: $\hat{H}^{(\text{d})}_{\text{free}}$, the free Hamiltonian of the detector, $\hat{H}^{(\text{f})}_{\text{free}}$, the free Hamiltonian of the field, and the field-detector interaction Hamiltonian $\hat{H}_{\text{int}}$:
\bea\label{Hamiltonian1}
\hat{H}(\tau)=\hat{H}^{(\text{d})}_{\text{free}}+\hat{H}^{(\text{f})}_{\text{free}}+\hat{H}_{\text{int}}(\tau).
\eea
We model the detector-field interaction with the well-known Unruh-DeWitt interaction \cite{DeWitt,Louko2008}
\bea
\hat{H}_{\text{int}}= \lambda\chi(\tau) \hat{\mu}(\tau) \hat{\phi}\left[x(\tau)\right],
\eea
where the constant $\lambda$ is the coupling strength, $\chi(\tau)$ is the {\it switching function} or {\it time window function} controlling the smoothness of switching the interaction on and off.  $\hat{\mu}(\tau)$ is the monopole moment of the detector and $\hat{\phi}\left[x(\tau)\right]$ is the massless scalar field to which the detector is coupling. We consider the coupling constant to be a small parameter so we can work with  perturbation theory to second order in $\lambda$. In our setting, the switching function is nonvanishing only during the time the atom spends in the cavity, i.e., $\chi(\tau)=1$ during $0\le \tau \le T$. The monopole moment operator of the detector has the usual form in the interaction picture \cite{Birrell1982,Brown2012,Fuenetesevolution},
\bea \label{monopole}
\hat{\mu}_{\text{d}}(\tau)=(\hat\sigma^{+} e^{\ii\Omega_{\text{d}}\tau}+\hat \sigma^{-}e^{-\ii\Omega_{\text{d}} \tau}),
\eea
in which, $\Omega_{\text{d}}$ is the proper energy gap between the ground state, $\left| g \right\rangle$ and the excited state, $\left| e \right\rangle$ of the detector and $\hat \sigma^{-}$ and $\hat \sigma^{+}$ are ladder operators.

Expanding the field  in terms of an orthonormal set of solutions inside the cavity yields the Hamiltonian in the interaction picture \cite{Brown2012}
\be\label{hamilto1}
\hat{H}_{\text{int}}(t)\!=\!\lambda \! \sum^{\infty}_{n=1}\!\frac{\hat{\mu}_{\text{d}}(\tau)}{\sqrt{\omega_n L}}\big(\hat{a}^{\dagger}_{n}u_n\left[x(\tau),t(\tau) \right]+\hat{a}_{n}u^{*}_n\left[x(\tau),t(\tau)\right]\!\big)
\ee
We will consider  Dirichlet (reflective) boundary conditions for our cavity,
\bea
\phi\left[0,t\right]=\phi \left[L,t\right]=0
\eea
and since we are in the Minkowski background, the field modes take the form of stationary waves
\bea
u_n\left[x(\tau),t(\tau)\right]=e^{\ii\omega_n t(\tau)}\sin \left[k_n x(\tau)\right],
\eea
where $\omega_{n}=\left|k_n\right|=n \pi/L$.


To characterize the vacuum response of the particle detector undergoing trajectory \eqref{traj1}, we initially prepare the detector in the ground state and the field in the optical cavity in a coherent state $\left|\alpha\right\rangle$. We choose the coherent state to be in the $j$-th cavity mode  with frequency $\omega_{j}=j \pi/L$, while the rest of the cavity modes are in the ground state. This way the main effects will not come from vacuum fluctuations but will instead be amplified by the stimulated emission and absorption of the atom coupled to the coherent state \cite{AasenPRL,Marvy2014}. Therefore the initial state of the system will be
\bea \label{initialC}
\rho_0=\proj{g}{g} \otimes \proj{\alpha_j}{\alpha_j}\bigotimes_{n\neq j} \proj{0_n}{0_n}.
\eea

While passing through the cavity, the detector spends a period of time T inside the cavity. Time evolution of the system is governed by the interaction Hamiltonian \eqref{hamilto1} in the proper frame of the detector. We define a time evolution operator for the detector inside the cavity to be
\be
\hat{U}(T,0)\!=\!\openone\underbrace{-\ii\!\int^{T}_{0}\!\!\!d\tau \hat{H}_{\text{int}}(\tau)}_{\hat{U}^{(1)}}\underbrace{-\!\!\!\int^{T}_{0}\!\!\!d\tau\!\! \int^{\tau}_{0}\!\!\!d\tau'\hat{H}_{\text{int}}(\tau)\hat{H}_{\text{int}}(\tau')}_{\hat{U}^{(2)}}+ ...
\ee

Therefore the system's density matrix at the time T would be evaluated as \cite{AasenPRL}
\be
\rho_{T}\!=\!\big[\openone+\hat{U}^{(1)}+\hat{U}^{(2)}+\mathcal{O}(\lambda^3)\big]\rho_0\big[\openone+\hat{U}^{(1)}+\hat{U}^{(2)}+\mathcal{O}(\lambda^3)\big]^{\dagger}.
\ee

Using the interaction Hamiltonian and the time evolution operator we defined above, the first order term of the perturbative expansion takes the following form
\begin{align} \label{evolution}
\hat{U}^{(1)}=\frac{\lambda}{\ii}\sum^{\infty}_{n=1}&\big[\sigma^{+} a^{\dagger}_{n}I_{+,n}+\sigma^{-} a_{n}I^{\ast}_{+,n}\nn\\
&+\sigma^{-} a^{\dagger}_{n}I_{-,n}+\sigma^{+}a_{n}I^{\ast}_{-,n}\big],
\end{align}
where $I_{\pm,n}$ is
\bea\label{integral}
I_{\pm,n}=\int^{T}_0 d\tau~e^{\ii\left[\pm \Omega_{\text d} \tau+\omega_nt(\tau)\right]} \sin\left[k_n x(\tau)\right],
\eea

To compute the density matrix for the detector, $\rho_{T}^{(\text{d})}$, we need to take the partial trace over the field degrees of freedom  \cite{AasenPRL}. The leading contribution comes from second order in the coupling strength, $\lambda$ and the final form of the detector density matrix will be \cite{Aida12014}
\begin{align}\label{densitytrace}
\rho_{T,(\text{d})}&=\text{Tr}_{(\text{f})}\!\left[\rho_{0}+\hat{U}^{(1)}\rho_0\hat{U}^{(1)\dagger}\!+\hat{U}^{(2)}\rho_0+\rho_0\hat{U}^{(2)\dagger}\right]\!\!,
\end{align}
which yields
\bea
\rho_{T,(\text{d})}=\text{Tr}_{\text{f}}\, \rho_T=\left[\begin{array}{ll}
1-P_{\alpha}&0\\
0&P_{\alpha}\end{array}\right].
\eea
$P_{\alpha}$ is the transition probability of the detector from the ground state to the first excited state to leading order in perturbation theory, given by \cite{Marvy2014}
\begin{equation} \label{PC}
P_{\alpha}(\epsilon,\gamma)=\frac{\lambda^2}{L}\left[\frac{\alpha^2}{k_{\alpha}}\big(\left|I_{+,j}\right|^2+\left|I_{-,j}\right|^2\big)+\sum_{n=1}^{\infty}\left|I_{+,n}\right|^2\right],
\end{equation}
where $\alpha$ is the amplitude of the coherent state. Notice that the probabilities $P_{\alpha}(\epsilon,\gamma)$ depend on $\gamma$ and $\epsilon$ through the integrals $I_{\pm,n}$, given in \eqref{integral} as functions of $x(\tau)$ and $t(\tau)$. $x(\tau)$ and $t(\tau)$ dependence on $a_0,\gamma,\epsilon$ is obtained by substituting  \eqref{speed} and \eqref{four} into \eqref{traj1}.

\subsection{Alignment metrology}

In the alignment metrology setting, we study the sensitivity of the response of a detector to small harmonic spatial perturbations of its otherwise constantly accelerated trajectory, and analyze its possible use as a witness of the relative alignment of an optical cavity with a beam of atomic detectors. In this setting, the atomic probes move along a constantly accelerated trajectory which undergoes a spatial perturbation that is harmonic in the cavity's reference frame, $(x,t)$:
\bea\label{traj2}
x(t)&=&\frac{1}{a}\left[\sqrt{1+a^2 t^2}-1 \right]+ \epsilon \sin(\gamma t)
\eea 
where $\epsilon$ and $\gamma$ are characterizing the amplitude and frequency of the perturbation, respectively. In this case, since the motion is analyzed from the lab's frame, we need to find the (rather non-trivial) relationship between the proper time of the accelerated atom and the cavity frame. The relationship between the cavity frame's proper time and the atomic probe's proper time can be worked out from
\bea
\left(\frac{d \tau}{d t}\right)^2=1-\left(\frac{d x}{d t}\right)^2.
\eea
Solving this differential equation for $d \tau/dt$ together with \eqref{traj2} yields
\begin{equation}\label{Arghh}\tau(t)=\frac{\text{arcsinh} (a t)}{a}-\frac{a \epsilon \Big(\cos(\gamma t)+t \gamma \sin(\gamma t)\Big)}{\gamma}+{\cal O}(\epsilon^2).
\end{equation}

\begin{figure}[htp]
	\includegraphics[width=0.2\textwidth]{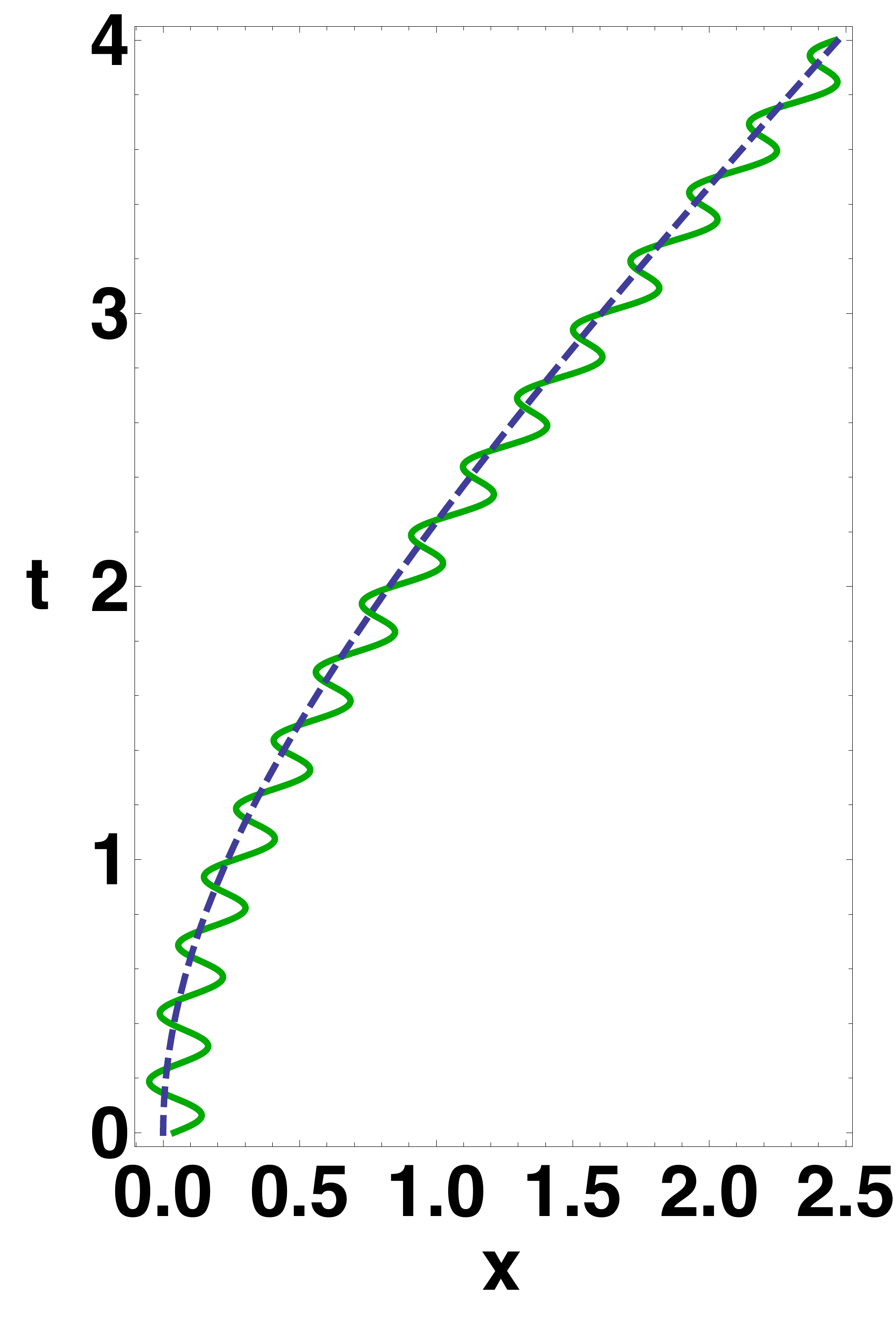}
  \caption{(Color online) The non-perturbed (blue-dashed curve) and perturbed (green-solid curve) trajectory for the alignment metrology detting. The trajectory is parameterized in the lab's frame $(x,t)$.}
  \label{alignment}
	\end{figure}

The general form of this trajectory is shown in Fig. \ref{alignment} for both the perturbed and  the non-perturbed cases. 

While crossing the cavity, the detector spends a period of time T  in traversing the full length along its trajectory. In order to find the time evolution of the system, we first need to find the form of the atom-field Hamiltonian that generates evolution for the entire system with respect to the time coordinate of the lab frame, $t$. The way to obtain this is explained in detail in \cite{Brown2012}.  The correct time-reparametrization of \eqref{Hamiltonian1} in terms of $t$ is given by
\bea\label{Hamiltonian2}
\hat{H}(t)=\frac{d\tau}{dt}\hat{H}^{(\text{d})}_{\text{free}}[\tau(t)]+\hat{H}^{(\text{f})}_{\text{free}}(t)+\frac{d\tau}{dt}\hat{H}_{\text{int}}[\tau(t)].
\eea
The monopole moment operator takes the usual form
\bea
\hat{\mu}_{\text{d}}(t)=\big(\sigma^{+}e^{\ii\Omega_{\text{d}} \tau(t)}+\sigma^{-}e^{-\ii\Omega_{\text{d}} \tau(t)}\big),
\eea
and interaction Hamiltonian becomes

\be \label{hamilto2}
\hat{H}_{\text{int}}(t)=\lambda \frac{d\tau}{dt}\sum^{\infty}_{n=1}\frac{\hat{\mu}_{\text{d}}(t)}{\sqrt{\omega_n L}}\big(\hat{a}^{\dagger}_{n}u_n\left[x(t),t\right]+\hat{a}_{n}u^{*}_n\left[x(t),t\right]\big),
\ee
with
\bea 
u_n\left[x(t),t\right]=e^{\ii\omega_n t}\sin \left[k_n x(t)\right].
\eea 	
Working in this frame,  the form of the function $I_{\pm,n}$ in the time evolution operator \eqref{evolution} turns into
\bea
I_{\pm,n}=\int^{T}_0 dt~e^{\ii\left[\pm \Omega_{\text d} \tau(t)+\omega_nt\right]} \sin\left[k_n x(t)\right].
\eea

Using the same approach as in the accelerometer setting, for characterizing the vacuum response of the particle detector undergoing trajectory \eqref{traj2}, we prepare a coherent state \eqref{initialC} for the scalar field to which the ground state of the detector is coupled and find the transition probabilities from \eqref{PC}.

\section{Results}\label{result}

The transition probabilities of atomic detectors crossing the cavity contain information about the parameters characterizing the detectors' motion. Of course, we would not want to use perturbations of the Unruh temperature as a means to characterize the the trajectory of the detector. This would be a rather futile endeavour since the the Unruh temperature itself is something extremely difficult to measure, let alone small perturbations of it. Instead, we will operate in a non-equillibrium regime where the detector will not have enough time to thermalize with the `modified' Unruh radiation. Therefore, we let the detector spend a small amount of time inside the cavity such that it does not thermalize with its environment. On top of that, and as discussed above, we consider a coherent state background which helps amplify the signal. This  is the reason why we may expect our system to show more sensitivity to the atom's trajectory. In this section, we analyze the sensitivity of the response of the detectors to perturbations in the kinematical parameters of the detectors' trajectory that we want to measure, both in the accelerometer and the alignment settings.

We pause to remark that our choice of switching function $\chi(t)$ (shown above equation \eqref{monopole})
removes the interaction between the field and the atom is off when the atom is outside of the cavity.   This assumption needs some justification since one cannot just `switch off' the interaction of the atom with the electromagnetic field when it is outside the cavity. The rationale of this assumption is twofold. On one hand we assume that the atomic state preparation happens at the entrance of the cavity, when the atom's speed is zero. Equivalently, we are considering a situation in which the atom is post-selected to be in its ground state prior to entering the cavity, and so pre-existing excitations as may be present outside of the cavity are not relevant. On the other hand, the main effects on the atomic state responsible for the results reported here are provoked by the variation of the boundary conditions and the perturbation of the atomic trajectory, which are amplified by the fact that the trajectory is relativistic. As we discussed above, the signature of the Unruh effect itself is small as compared to the  non-equilibrium effects coming from the time dependence of the trajectory perturbations. Therefore if the flight of the atom includes some small segments of free flight (outside the cavity), since the Unruh noise would be in these cases arguably negligible it should not modify our results. 

\subsection{A quantum accelerometer}

We focus first on the accelerometer setting, in which there might be  small fluctuations of the probe's acceleration of the detector in its own proper frame.  We will model this by 
 assuming that the proper acceleration 
of a set of uniformly accelerated detectors is perturbed by a small harmonic function.   One possible way to think about these time dependent oscillations is associate them with possible  inexactnesses in the measure of the acceleration in the proper frame of the detector, so that through relativistic quantum effects we may expect to be able to use the internal degree of freedom of the atomic probe to increase the accuracy in exactly determining this proper acceleration. 

With this aim, we study the sensitivity of the transition probability of the detector to the amplitude of the harmonic perturbations and characterize the spectral response of the setting to the specific frequency range of the perturbations. The detector's trajectory (with a harmonically perturbed acceleration) is given by inserting \eqref{four} in \eqref{traj1}.

To study how sensitive the setting is to the parameters of the perturbation, we will analyze the following sensitivity estimator:

\begin{equation}\label{estim}S(\epsilon,\gamma)=\frac{|P_{\alpha}(\epsilon,\gamma)-P_{\alpha}(0,\gamma_0)|}{P_{\alpha}(0,\gamma_0)}\end{equation}
where $P_{\alpha}(\epsilon,\gamma)$ is the transition probability of the detector with a perturbed acceleration given by  \eqref{four}, and $P_{\alpha}(0,\gamma_0)$ is the transition probability for a constantly accelerated detector whose trajectory is unperturbed.

\begin{figure}[htp]
   \includegraphics[width=0.48\textwidth]{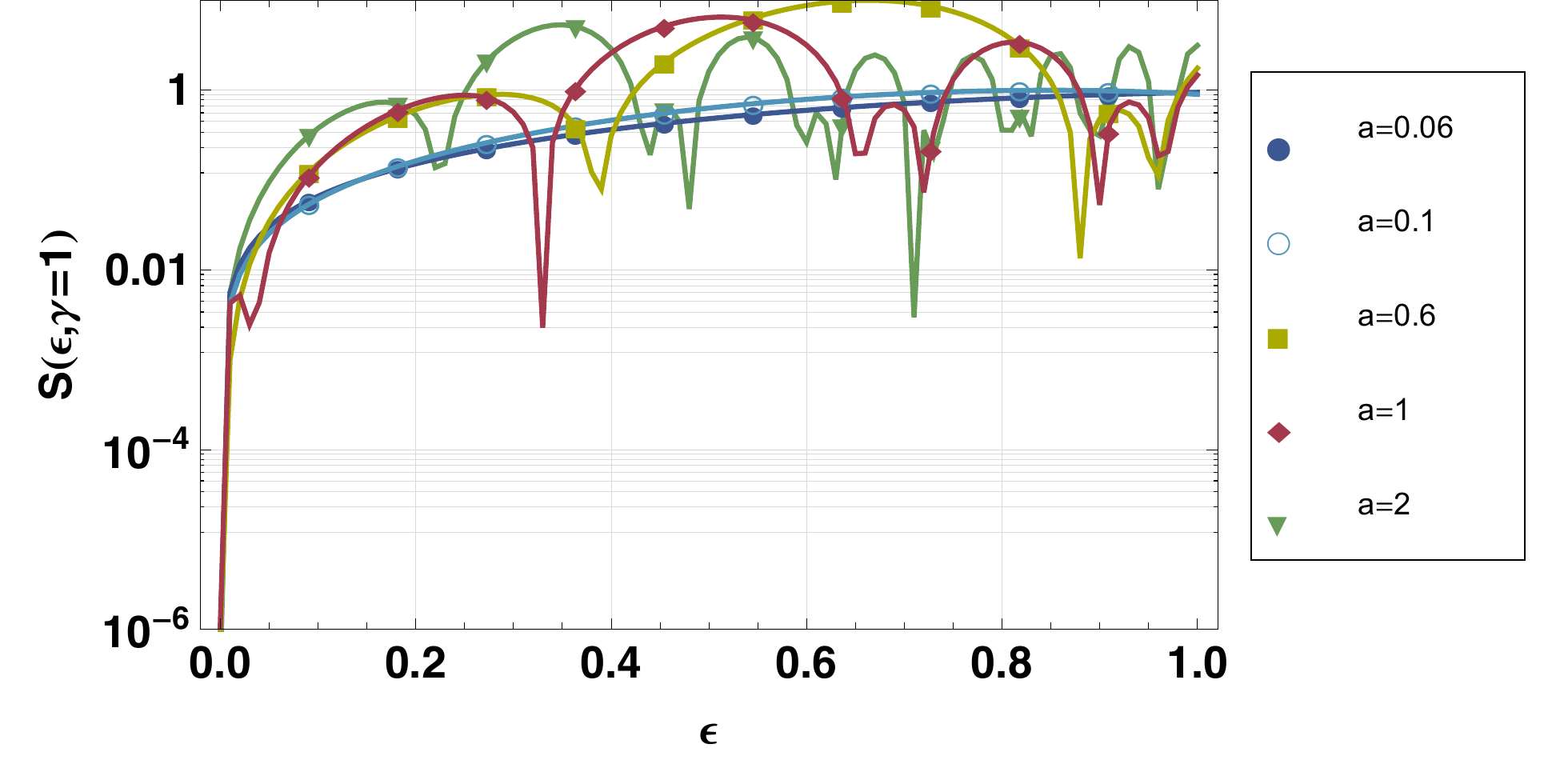}
    \caption{(Color online) Behaviour of the sensitivity of the detector's transition probability as a function of the amplitude of a perturbed proper acceleration for different initial accelerations.}
   \label{accelerometeramp}
\end{figure}

\begin{figure}[htp]
\includegraphics[width=0.48\textwidth]{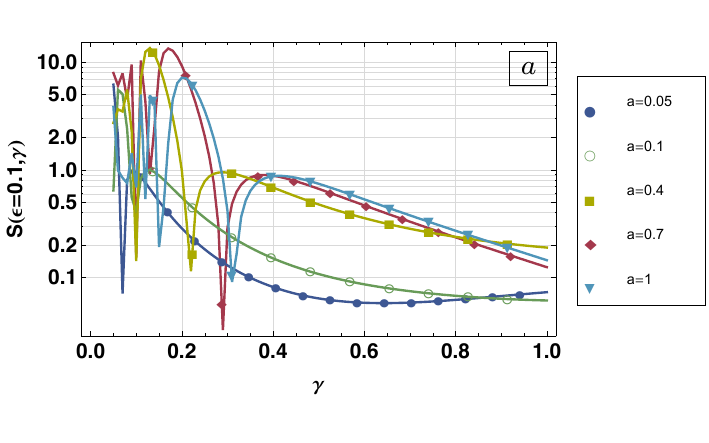}
\includegraphics[width=0.48\textwidth]{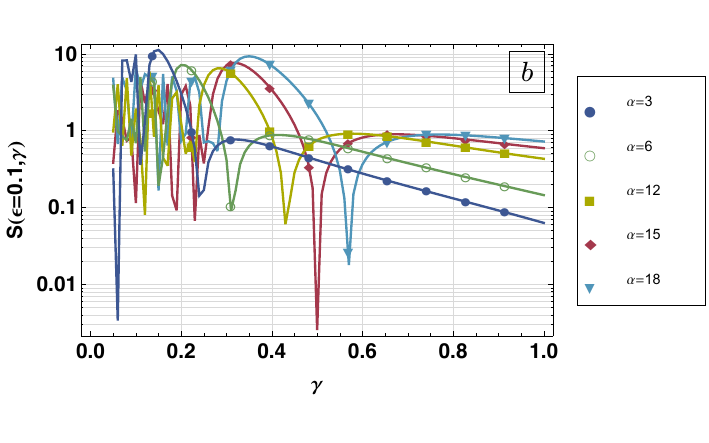}
\includegraphics[width=0.48\textwidth]{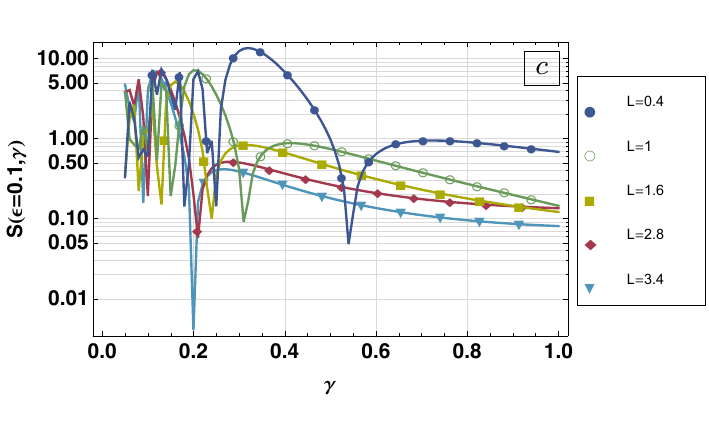}
\caption{(Color online) Spectral response of the detector for a) both relativistic and nonrelativistic accelerations, b) different modes of the field which are in coherent states and coupled to the ground state of the detector and c) different lengths of the cavity.}
   \label{accelerometerplots}
\end{figure}

Fig. \ref{accelerometeramp} shows the explicit dependence of the sensitivity estimator \eqref{estim} on the parameters characterizing the perturbation,. Namely, it shows the sensitivity of the response of the detector to the amplitude $\epsilon$ of the perturbations for different values of acceleration, whereas  the spectral response of the sensitivity to different values of the perturbation frequency ($\gamma$) is shown in Fig. \ref{accelerometerplots}. 

As one can observe in Fig. \ref{accelerometeramp}, for small accelerations,  closer to the regimes where the atom does not attain relativistic speeds while crossing the cavity, the sensitivity (to acceleration perturbations) of the detector's transition probability is monotonic on the amplitude of the perturbations. However, for large accelerations the sensitivity does not behave monotonically, and there appear specific amplitudes for which the sensitivity dips. The spectral response displayed in Fig. \ref{accelerometerplots} shows that the response of the detector is always more sensitive to the lower frequencies. The behaviour for higher frequencies depends on the energy gap of the atomic probe. For a fixed gap, the sensitivity of the probe seems to be exponentially suppressed as the frequency of the perturbations grows. One possible way to understand this is that when the frequency of the harmonic acceleration perturbation is much higher than the frequency associated with the transition of the atom, the the atomic probe is primarily responsive to its average constant acceleration; the perturbations are much faster than the dynamics of the atom and so become invisible to it. However, as we see in Fig. \ref{accelerometerplots}b),  it is possible to adjust the gap of the atomic transition used as a probe to tune out to a specific frequency range of the perturbations. 

In Fig. \ref{accelerometerplots}c), we show how sensitive the response of the atomic probes is to the length of the cavity they're transversing. This in turns also determines how much relativistic the probes are when existing the cavity for constant acceleration. These curves also suggest that it may be possible to use similar settings as a means to determine the length of an optical cavity.

Of course, the sensitivity estimator we studied only gives us an idea of the potentiality of these settings for the measurement of the parameters of the perturbation. A more realistic practical implementation of such settings would require  considerable effort. For example this might be implemented by comparing one setting where all the parameters are known with another setting where the parameters are not known. The comparison of the transition rates of beams of atoms in these two settings may reveal the information about the parameters to be determined. In such a comparison the estimator built here becomes relevant.

\subsection{Alignment metrology}

In the alignment setting we assume that the trajectory of uniformly accelerated detectors is perturbed by a small harmonic motion, that we could, for instance, ascribe to oscillations of the trajectory of the detector in the cavity frame. These can be understood as time dependent imprecisions in the alignment of the setting with the optical cavity.

\begin{figure}[htp]
   \includegraphics[width=0.48\textwidth]{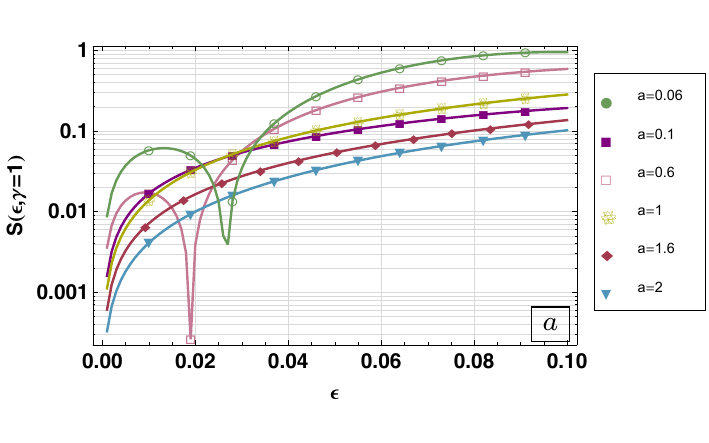}
	\includegraphics[width=0.48\textwidth]{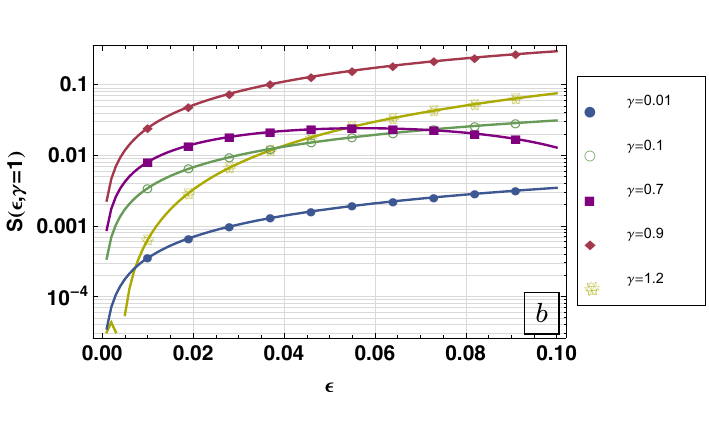}
    \caption{(Color online) The sensitivity of the excitation probability of the detector to the amplitude $\epsilon$ of the trajectory perturbations for a) different constant accelerations and for b) different frequencies of perturbation.}
   \label{alignmentamp}
\end{figure}

Here we study the sensitivity of the detector's response to the amplitude of the harmonic perturbations and characterize the spectral response of the setting to the frequency of perturbations. We consider the spatial perturbation as expressed in equation \eqref{traj2}. Since in the derivation of the parametrization of the detector's world line \eqref{Arghh} we linearized in the amplitude of the perturbation $\epsilon$, we only consider small amplitudes $0<\epsilon<0.1$ in our study. The sensitivity of the response of the detector as a function of the amplitude $\epsilon$ for different values of acceleration and different frequencies are shown in Fig. \ref{alignmentamp}a) and b) respectively. We estimate this sensitivity by using the same quantity \eqref{estim} as in the accelerometer setting with the only difference that $P_{\alpha}(\epsilon,\gamma)$ represents transition probability of the detector with a spatially perturbed trajectory which is otherwise constantly accelerated. 

As shown in Fig. \ref{alignmentamp}a) for small accelerations where the system is closer to nonrelativistic regimes, the detector's response shows more sensitivity to the perturbation of its trajectory than in the case of higher accelerations (relativistic regimes). In contrast to  the previous case of perturbations in the probe's proper acceleration, we see from Fig. \ref{alignmentamp}b)  that the detector's response is less sensitive to low frequency perturbations of its spatial trajectory. This is again reasonable, considering that higher the frequency of perturbations of the spatial trajectory in the lab frame, the more of an effective change they will have on the detector's proper acceleration; a high frequency spatial perturbation in the lab frame corresponds to a large instantaneous change of the proper acceleration of the detector. This in turn affects the response of the detector more dramatically than if the perturbation of the spatial trajectory is slow.  As expected, the sensitivity increases monotonically as the amplitude of fluctuations grows, as  seen in the figures.

\begin{figure}[htp]
   \includegraphics[width=0.48\textwidth]{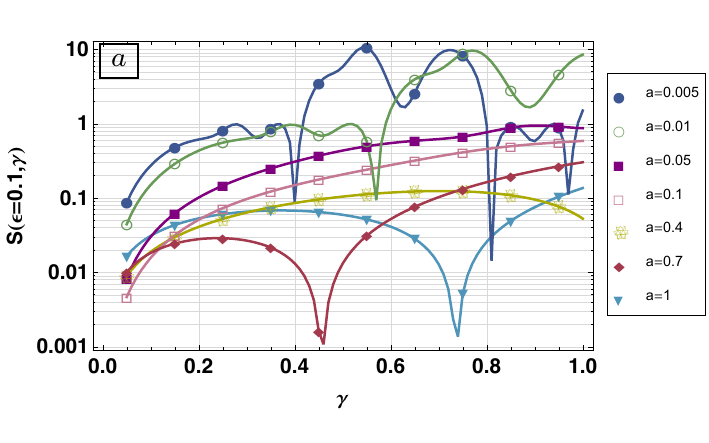}
\includegraphics[width=0.48\textwidth]{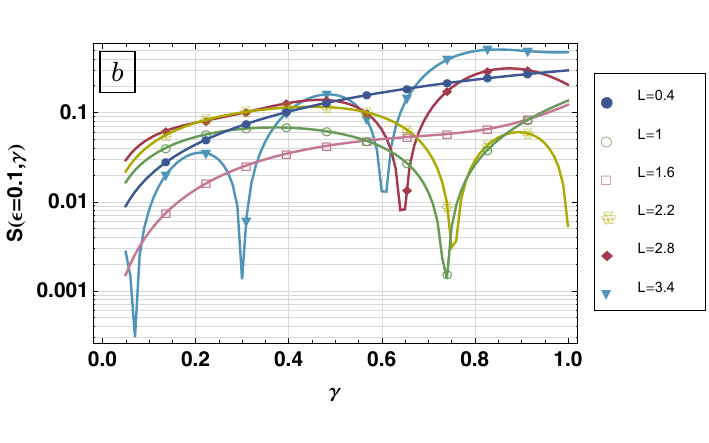}
\includegraphics[width=0.48\textwidth]{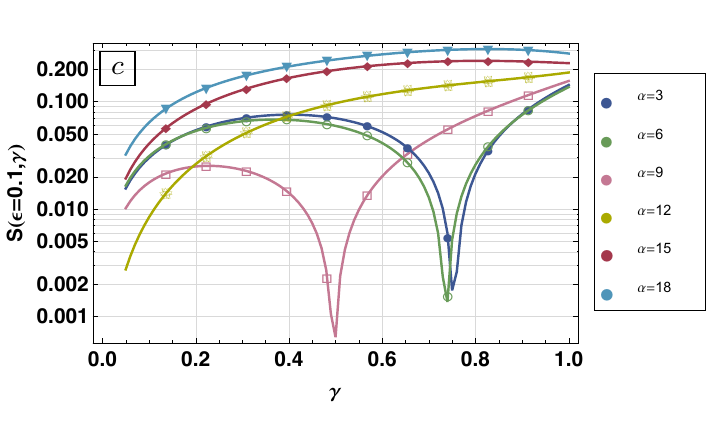}
    \caption{(Color online) Spectral response of the detector for a) different accelerations from nonrelativistic regimes $(a=0.005, 0.01, 0.05)$ to relativistic regimes $(a=0.1, 0.4, 0.7, 1)$, b) different lengths of the optical cavity and c) different modes of the field which are in coherent states and coupled to the state of the detector.}
   \label{alignmentplots}
\end{figure}

We display in Fig. \ref{alignmentplots}  the spectral response of the sensitivity of the probe's excitation probability for a fixed amplitude of the perturbation for different values of the setting parameters:   proper accelerations Fig. \ref{alignmentplots}a),  cavity lengths Fig. \ref{alignmentplots}b) and  detector gaps Fig. \ref{alignmentplots}c).

The general trend in all cases is that the transition probability of the detector presents dips for specific values of the perturbation frequency $\gamma$. In other words, there are some specific perturbation frequencies for which the sensitivity of the setting goes down abruptly, being the position of these dips is a function of the system parameters. This resonance-like effect that may be related with the spatial distribution of the cavity modes as seen from the reference frame of the atom whose trajectory is perturbed, but it seems to depend non-trivially on the system parameters and we have not been able to identify its exact origin through numerical analysis.

\section{Conclusions}\label{concl}

We have analyzed the sensitivity of the response of a constantly accelerated atomic probe, transversing an optical cavity, when its trajectory is perturbed. We showed that the probe's transition probability is,   in principle,  sensitive to small deviations from constant acceleration. We conclude that the transition rate of a beam of atoms transversing optical cavity can provide information about its past spatial trajectory.

We have theoretically studied the potential of the use of an atomic internal quantum degree of freedom  to design novel quantum metrology settings. In particular we considered two scenarios: one where the probe undergoes small time-dependent perturbations of its proper acceleration, and another one when the probe's trajectory experiences small spatial time-dependent perturbations as seen from the laboratory's frame. 

The first scenario could correspond to an accelerometer setting where we use the internal degree of freedom of the atom to identify small time-dependent forces acting on the probe that will cause it to deviate from  constant proper acceleration. The second scenario could correspond to an alignment measurement setting where we use the internal atomic degree of freedom to characterize small vibrations or imperfections of the alignment of an optical cavity with a beam of atoms transversing it. 

While an analysis of a proper experimental implementation goes beyond the scope of this paper, these findings have a potential use in quantum metrology of optical setups. For instance one could  compare one setting where all the parameters are known with another setting where they  are not known. 
 In practice, however, the  ratio of the probabilities will be subject to significant statistical fluctuations that could  mask the effects we have obtained. To achieve the sensitivity levels that are potentially available, the implementation of our scheme will require accumulation of statistics over a number of identical experiments by sending a large number of atomic probes through the cavity. Thus, by analyzing the transition rates of different atomic beams, one could  in principle deduce the specific form of the trajectory of such beams or infer the parameters of the optical cavities they are traversing. 

\section{Acknowledgments}

E.M-M. acknowledges the support of the Banting Postdoctoral Fellowship Programme. This work was supported in part by the Natural Sciences and Engineering Research Council of Canada.

\bibliography{cavity_refs}

\end{document}